\documentclass[fleqn,12pt,twoside]{article}
\usepackage{espcrc1}

\usepackage{graphicx}
\usepackage[figuresright]{rotating}

\newcommand{\AmS}{{\protect\the\textfont2
  A\kern-.1667em\lower.5ex\hbox{M}\kern-.125emS}}

\title{Chiral Expansion, Renormalization and the Nuclear Force}

\author{\underline{E. Ruiz Arriola} \address[MCSD]{Departamento de F{\'{\i}}sica At\'omica, Molecular y Nuclear (Granada) Spain}
\thanks{Speaker at International IUPAP Conference on Few-Body Problems
in Physics (FB18), Santos-Sao Paulo (Brasil), 21-26 August 2006.  Work
supported by DGI and FEDER funds, under contract FIS2004- and by the
Junta de Andaluc\'\i a grant no. FM-225 and EURIDICE grant number
HPRN-CT-2003-00311.} and M. Pav\'on Valderrama\addressmark}
       
\begin{document}


\maketitle       

\begin{abstract}
The renormalization of singular chiral potentials as applied to NN
scattering and the structure of the deuteron is discussed. It is shown
how zero range theories may be implemented non-perturbatively as
constrained from known long range NN forces.
\end{abstract}

\section{Introduction}

A piece of standard wisdom in nuclear physics has been the long
distance dominance of pion exchanges of the nuclear force while the
short range component is fairly
unknown~\cite{deSwart:1995ui}. 
Such a situation seems a fertile ground for Effective Field
Theories~\cite{Bedaque:2002mn,Epelbaum:2005pn}.  Actually, the
functional dependence of the long range piece is calculable within
chiral perturbation theory~\cite{Kaiser:1997mw,Rentmeester:1999vw}
yielding for the reduced potential ($U (r) = M V(r)$)
\begin{eqnarray}
U (r) =  M_N m_\pi \Big\{ \frac{m_\pi^2}{f_\pi^2} W^{(1)} (m_\pi r ) + \frac{m_\pi ^4}{f_\pi^4}
W^{(2)} (m_\pi r ) + \frac{m_\pi^4}{f_\pi^4} \frac{m_\pi}{M_N}
W^{(3)} (m_\pi r ) + \dots \Big\} \, \quad (r \neq 0) 
\label{eq:pot_chpt}
\end{eqnarray}
with $M_N$ the nucleon mass and $m_\pi $ the pion mass and $f_\pi$ the
pion weak decay constant. Based on these or similar chiral
potentials~\cite{Entem:2003ft,Epelbaum:2004fk} successful description
of low energy scattering data~\cite{deSwart:1995ui} is achieved.
However, one of the problems which immediately arises refers to the
location of a sensible boundary between short and long distances such
that model independence is ensured. The most obvious candidate for a
short distance cut-off is the smallest probing wavelength, which in
the case of elastic NN interaction corresponds to $\lambda = 0.5 {\rm
fm}$. Thus, to describe satisfactorily elastic NN scattering from
threshold up to the pion production threshold a fairly {\it good
knowledge} of the force above $0.5 {\rm fm}$ is needed. The crucial
question is whether such a statement can be verified regardless on the
uncertainties at short distances. Any effective description of long
wavelength phenomena requires that distances shorter than the probing
wavelength become irrelevant, and although the statement is quite
obvious as a physical requirement the corresponding mathematical
implementation is far from trivial. The local character of
Eq.~(\ref{eq:pot_chpt}) suggests using coordinate space methods which,
due to the boundary value character of the Schr\"odinger equation have
the obvious advantage of a rather neat separation between disjoint
regions of space which can be connected at the boundaries. This
property does not hold in momentum space, where all scales are
intertwined. In addition, a clear advantage of this coordinate space
analysis is the hierarchy of equations that arises in the
renormalization
problem~\cite{PavonValderrama:2003np,PavonValderrama:2004nb}.


\section{The singlet $^1S_0$channel}

Our points are best illustrated for s-waves ($^1S_0$) where the
Schr\"odinger equation reads
\begin{eqnarray}
-u''_k (r) + U (r) u_k(r)  = k^2 u_k (r) 
\end{eqnarray} 
The asymptotic solution for finite energy is normalized as
\begin{eqnarray}
u_k (r) \to \frac{\sin ( k r + \delta_0)}{\sin \delta_0 } \quad \left( r \to \infty \right) 
\label{eq:delta0}
\end{eqnarray} 
In the zero energy limit (using $\delta_0 (k) \to -\alpha_0 k +
\dots$) one gets
\begin{eqnarray}
u_0 (r) &\to& 1 - \frac{r}{\alpha_0} \qquad ( r \to \infty )
\label{eq:alpha0}
\end{eqnarray} 
Orthogonality of solutions $0 \le r \le \infty $ implies that 
\begin{eqnarray} 
\delta (k-p) = N \int_0^\infty u_k (r) u_{p} (r) dr 
\end{eqnarray} 
with $N$ a suitable normalization constant. Straightforward
manipulation yields
\begin{eqnarray} 
0 = u_k' u_p - u_k u_p' \Big|_{0^+}  
\label{eq:orth}
\end{eqnarray} 
which implies an {\it energy independent boundary condition}. In
general a short distance limit $r_c \to 0^+$ may be required.  The
relation to a sharp cut-off in momentum space is $r_c= \pi / 2
\Lambda$~\cite{PavonValderrama:2004td}. From a mathematical viewpoint
this is equivalent to looking for self-adjoint extensions of hermitian
operators on the Hilbert space with a {\it common domain} within which
completeness of solutions is ensured. This condition can be deduced
from the smallness of probability at small
distances~\cite{Valderrama:2005wv}.

In the absence of a potential, $U(r)=0$, (pionless theory),
Eqs.~(\ref{eq:delta0}) and (\ref{eq:alpha0}) become the solution
everywhere and hence taking the limit $p \to 0$ one has
\begin{eqnarray} 
k \cot \delta_0 (k)= \frac{u_k'(0)}{u_k(0)} = \frac{u_0'(0)}{u_0(0)} =
-\frac1{\alpha_0}
\end{eqnarray} 
This is the effective range expansion $ k \cot \delta_0 = -
\frac{1}{\alpha_0} + \frac12 r_0 k^2 + v_2 k^4 + \dots $ with $ r_0 =
v_2 = \dots =0 $. The vanishing $r_0$ is a sufficient condition for
causality, $r_0 \le 0$~\cite{Phillips:1996ae}.  Under a weak potential
perturbation one has, after renormalization, the result
\begin{eqnarray}
k \cot \delta_0 (k) =-\frac1{\alpha_0} + \int_{0}^\infty dr U(r) \left(
\left[ \cos(kr )  -\frac{\sin(k r)}{\alpha_0 k} \right]^2 - \left[
  1-\frac{r}{\alpha_0} \right]^2 \right) + \dots 
\end{eqnarray}
The renormalized effective range is {\it entirely predicted} from the
potential at all distances
\begin{eqnarray}
r_0 = 4 \int_{0}^\infty dr r^2 U(r) \left( 1-
\frac{r}{\alpha_0}\right)^2 + \dots
\label{eq:r0_pert} 
\end{eqnarray}
Note the extra power suppression at the origin when $\alpha_0$ is
fixed, indicating short distances become {\it less} important. Finite
cut-off approaches not only fix a short distance finite cut-off in the
lower limit $r_c$ but also {\it add} an extra short distance
contribution $r_{0,S}$. Also, if $r_{0,S} \neq 0 $ for $r_c =0$ there
is no {\it additional} power suppression of the potential at the
origin, which is counterintuitive. Orthogonality implies that
$r_{0,S}\to 0 $ as $r_c \to 0$, and similarly with the short distance
components of $v_2$, etc. The question is whether for potentials in
Eq.~(\ref{eq:pot_chpt}) the experimental $r_0$ can be entirely
saturated when the cut-off is removed.

\begin{figure}[t] 
 \begin{minipage}{3in}
   \includegraphics[width=3.0in,angle=0]{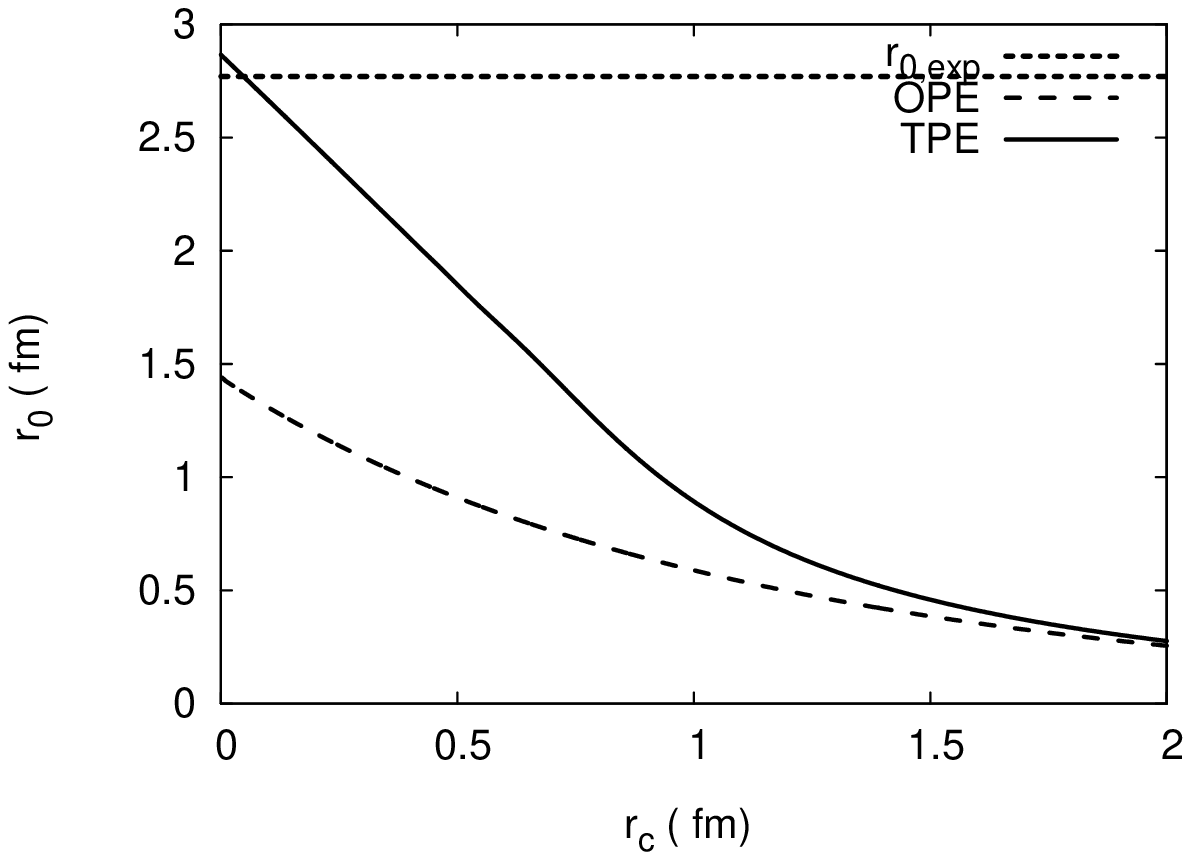}
 \end{minipage}
 \hfill
 \begin{minipage}{3in}
   \includegraphics*[width=3.0in]{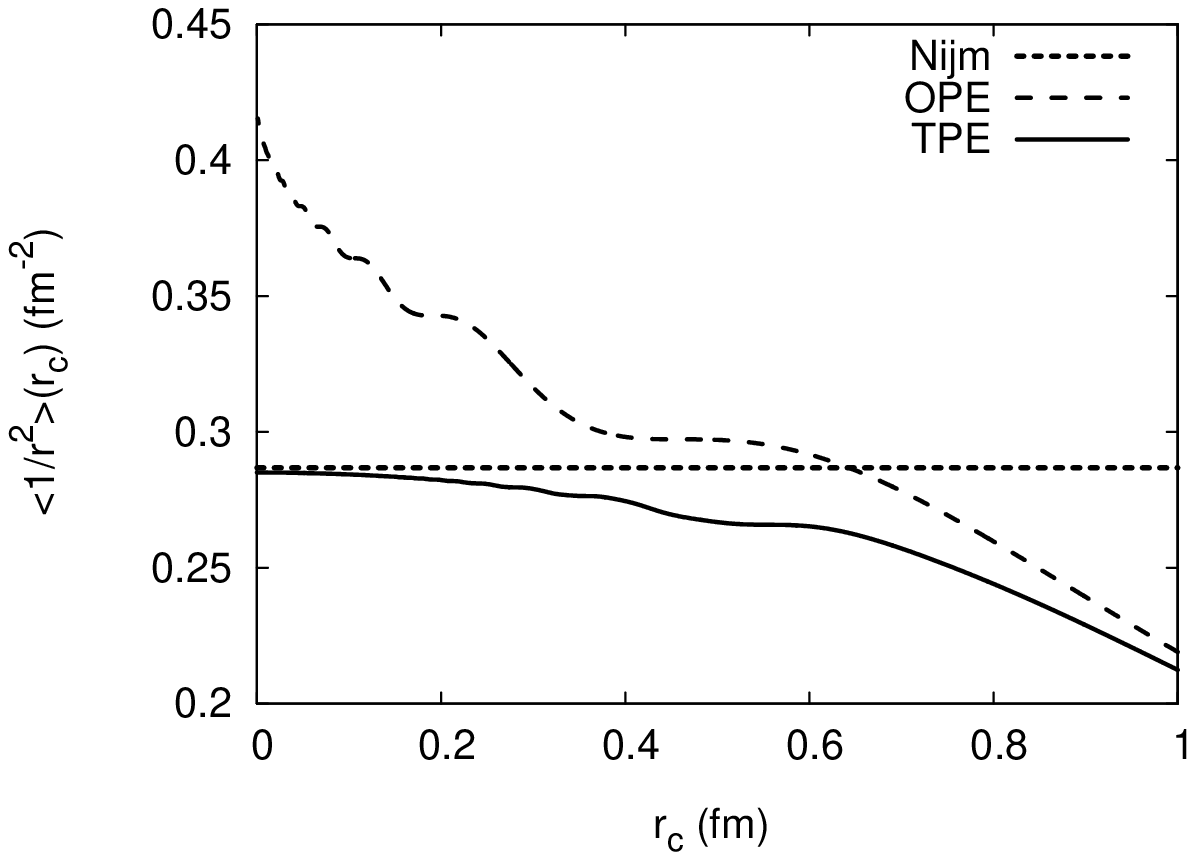}
 \end{minipage}
 \vspace*{-.3in}
 \caption{Cut-off dependence of OPE$_{\rm LO}$ and TPE$_{\rm N2LO}$
observables compared to the Nijm II and Reid 93 potential results.
Left: $^1S_0$ Effective range. Right: Deuteron $\langle r^{-2} \rangle
$ moment.}
\label{fig:one}
\end{figure}
Chiral potentials, although exponentially suppressed at large
distances develop power-like singularities at short distances, $  R^2 U(r) \to
- (R/r )^n $ with $R$ the typical short distance scale and $n \ge 
3$. For these potentials perturbation theory and orthogonality are
clearly incompatible if the cut-off is removed (see e.g. 
Eq.~(\ref{eq:r0_pert})). Non-perturbatively, however, there is no
problem. Indeed, the short distance behaviour of the wave function is
\begin{eqnarray}
u_k (r) &\to & C \left(\frac{r}{R}\right)^{n/4}
\sin\left[ \frac{2}{n-2} \left(\frac{R}{r}\right)^{\frac{n}2-1} +
\varphi \right] \quad ( r \to 0) \, .\label{eq:phi0} 
\end{eqnarray} 
The short distance phase $\varphi$ must be energy and potential
independent because of orthogonality. 
\begin{table}[tbc]  
\caption{$^1S_0$ threshold parameters, $ k \cot \delta_0 = -
\frac{1}{\alpha_0} + \frac12 r_0 k^2 + v_2 k^4 + \dots $, for fully
iterated potentials (f.i.) and for perturbation theory based on OPE distorted
waves~\cite{Valderrama:2005wv}. NijmII from \cite{deSwart:1995ui}.}
\begin{tabular}{|c|ccc|c|c|ccc|}
\hline $^1S_0 $ & LO$_{\rm f.i.}$ & NLO$_{\rm f.i.}$ & NNLO$_{\rm f.i.}$ & Exp.  & Nijm II & LO$_{\rm pert}$ & NLO$_{\rm pert}$ & N$^2$LO$_{\rm pert}$ \\ \hline
$\alpha_0( {\rm fm}) $ & Input & Input  & Input  & -23.74(2) & -23.73 & Input     &  Input & Input \\ 
$r_0( {\rm fm}) $ & 1.44 & 2.29  & 2.86  & 2.77(5) & 2.67 & 1.383     &  Input & Input \\ 
$v_2 ( {\rm fm}^3) $ & -2.11 & -1.02  & -0.36  & -- & -0.48 & -2.053 & Input  &  Input \\ 
$v_3 ( {\rm fm}^5) $ & 9.48 & 6.09  & 4.86  & -- & 3.96 & 9.484 & Input  & Input\\ 
$v_4 ( {\rm fm}^7) $& -51.31 & -35.16  & -27.64  & -- & -19.88 & -50.74 & -61.19  & -64.07  \\ 
\hline 
\end{tabular}
\label{tab:1}
\end{table}
The steps to follow are straightforward. i) For $k=0$ we fix
$\alpha_0$ to the physical value. ii) We integrate in
Eq.~(\ref{eq:alpha0}) and obtain the short distance phase $\varphi_0 $
from Eq.~(\ref{eq:phi0} ) iii) For $ k \neq 0 $ we use
$\varphi_k=\varphi_0 $ (orthogonality constraint from
Eq.~(\ref{eq:orth})) and iv) We integrate out and obtain $\delta_0$ from
Eq.~(\ref{eq:delta0}).  With this procedure we {\it predict} $
\delta_0 (k) $ from the potential and the scattering length $\alpha_0$
as {\it independent} quantities. An outstanding result is the
universal low energy theorem for the effective range, where the
potential and scattering length dependences can be disentangled
analytically, due to the superposition principle,
\begin{eqnarray}
r_0 = 2 \int_0^\infty dr ( 1 - u_{0,c}^2 ) -\frac{4}{\alpha_0}
\int_0^\infty dr ( r - u_{0,c} u_{0,s} ) + \frac{2}{\alpha_0^2} \int_0^\infty
dr ( r^2 - u_{0,s}^2 ) \, ,
\end{eqnarray} 
where $u_{0,c} (r) \to 1 $ and $ u_{0,s} (r) \to r $ for large $r$ are
zero energy solutions (see Ref.~\cite{Valderrama:2005wv} for more
results of this sort and further details). Numerically we find
(everything in {\rm fm} )
\begin{eqnarray} 
r_0 &=& 1.308 -
\frac{4.548}{\alpha_0}+\frac{5.193}{\alpha_0^2} \, \qquad ({\rm
LO} ) \nonumber \\ 
r_0 &= & 2.122 - \frac{4.889}{\alpha_0} + \frac{5.499}{\alpha_0^2} \, \qquad ({\rm
NLO} ) \, , 
\nonumber \\ 
r_0 &=& 2.672 -
\frac{5.755}{\alpha_0}+\frac{6.031}{\alpha_0^2} \, \qquad ({\rm NNLO}) 
\end{eqnarray} 
The cut-off dependence of the effective range can be seen in
Fig.~\ref{fig:one} for OPE and TPE. As we see in the NNLO case $r_0$
nicely approaches the experimental value as the cut-off is removed
$r_c \to 0 $, supporting the orthogonality constraints and the fact
the all contributions of the effective range can be deduced from a
good knowledge of the potential in a long distance expansion and 
model independent fashion. In table~\ref{tab:1} we compare results for
fully iterated potentials which need only one counterterm with a
distorted waves perturbation theory on the leading OPE interaction
where there is a proliferation of counterterms without any
substantial improvement.
The extension to
peripheral waves with $j \le 5 $ has been carried out in
Ref.~\cite{PavonValderrama:2005uj} for OPE (reproducing
\cite{Nogga:2005hy}) and non-perturbative TPE.
\begin{table}
\caption{\label{tab:table3} Deuteron properties and low energy
parameters in the $^3S_1-^3D_1$ channel for pionless, $U=0$, the
OPE and the TPE potential with $ \gamma= \sqrt{ 2 \mu_{np} B} $ with
$B=2.224575(9)$. Errors in TPE reflect the uncertainties in $\gamma$, $\eta$ and $\alpha_0$ only.
NijmII and Reid93 from \cite{deSwart:1995ui}.}
\begin{tabular}{|cccccccc|}
\hline  & $\gamma ({\rm fm}^{-1})$ & $\eta$ & $A_S ( {\rm
fm}^{-1/2}) $ & $r_m ({\rm fm})$ & $Q_d ( {\rm fm}^2) $ & $ \alpha_0
({\rm fm})$ & $r_0 ({\rm fm})$ \\ \hline
$U=0$ & Input & 0 & 0.6806 &  1.5265& 0  & 4.3177 & 0  \\ 
{\rm OPE}  & Input & 0.02633 & 0.8681 & 1.9351 & 0.2762 & 5.335 & 1.638 \\ 
TPE & Input & Input & 0.884(4) & 1.967(6) & 0.276(3)
& Input  & 1.76(3)\\ \hline \hline 
NijmII  & 0.23160 & 0.02521 & 0.8845(8) & 1.9675 & 0.2707 & 
 5.418  & 1.753 \\
Reid93 & 0.23160 & 0.02514 & 0.8845(8) & 1.9686 & 0.2703 & 
 5.422 & 1.755 \\ \hline 
Exp.  &  0.23160 &  0.0256(4)  & 0.8846(9)  & 1.971(6)  &
0.2859(3) & 5.419(7) &  1.753(8) \\ \hline 
\end{tabular}
\end{table}

\section{The deuteron}
The analysis for the deuteron equations has been carried out for OPE
and TPE in Refs.~\cite{PavonValderrama:2005gu} and
\cite{Valderrama:2005wv} respectively. Some results are compiled in
Table~\ref{tab:table3}. As we see they are rather good as compared
with realistic potential models. Actually, at NNLO
predictive power is lost, i.e. theoretical predictions have {\it
larger} error bars than experimental uncertainties. So, there is a
lack of motivation for doing N$^3$LO.  Moreover, the chiral constants are
determined from $^1S_0$, $^3S_1-^3D_1$ scattering data and the
deuteron yielding $ c_1 = -1.2 \pm 0.2 $, $ c_3 = -2.6 \pm 0.1 $ and
$c_4 = + 3.3 \pm 0.1 $ (in units of ${\rm GeV}^{-1}$). 

We have also analyzed $\pi d$ scattering at threshold in the fixed center
approximation~\cite{Valderrama:2006np}.  The multiple scattering
series yields (recoil and binding are neglected) 
\begin{eqnarray}
 a_{\pi d} = 2\Big[ b_0 + (b_0^2 - 2 b_1^2)
 \Big\langle r^{-1} \Big\rangle + (b_0^3 - 2 b_1^2
 b_0 - 2b_1^3 ) \Big\langle r^{-2} \Big\rangle + 
 (b_0^4 - 4 b_1^2 b_0^2 + 2 b_1^4 ) \Big\langle r^{-3} 
 \Big\rangle + \dots  \Big] \nonumber 
 \end{eqnarray} 
where $b_0$ and $b_1$ are isoscalar and isovector $\pi N$ scattering
lengths .  These inverse moments become convergent {\it precisely}
because the potential is iterated to all orders. The cut-off
dependence for $\langle r^{-2} \rangle $ can be seen in
Fig.~\ref{fig:one} for OPE and TPE and remarkable agreement for TPE as
$r_c \to 0 $ with the NijmII results is observed despite
the short distance enhancement. These moments diverge in perturbation
theory on boundary conditions or distorted OPE waves. $\langle r^{-3}\rangle $ is convergent for TPE and divergent for Nijm II and Reid93.

\section{Summary}

Perturbative treatments are systematic but require many more
counterterms than fully iterated potentials when the UV cut-off is
removed. On the other hand, renormalized non-perturbative approaches
suffer from a lack of systematics {\it a priori} in the sense that
strict dimensional power counting does not hold; corrections are
parametrically but non-analytically small and the number of
counterterms depends crucially on the attractive/repulsive character
of the chiral potential at short distances. Within this context we
find that Weinberg's power counting is incompatible with
renormalization for OPE (confirming \cite{Nogga:2005hy}) and TPE,
perhaps due to an incomplete inclusion of {\it all} TPE effects which
have similar range. Our approach focuses more on model independent
long distance correlations, not necessarily embodied by dimensional
power counting. Among those, we find central phases from TPE to be
largely explained just in terms of their pure Van der Waals components
whose coefficients depend on the chiral constants, suggesting a
unexpected connection to the liquid drop model from non-perturbative
renormalized chiral dynamics.

\end{document}